\documentclass[final,3p,times]{elsarticle}
\usepackage{graphicx}

\usepackage{geometry}
\newcommand{\dfrac}[2]{\displaystyle{\frac{#1}{#2}}}
\newcommand{\diff}[2]{\frac{\partial #1}{\partial #2}}
\newcommand{\mapright}[1]{\smash{\mathop{\hbox to 1.5cm{\rightarrowfill}}\limits^{#1}}}
\newcommand{\lw}[1]{\smash{\lower 1.5ex\hbox{#1}}}

\journal{Physica A}

\begin{document}

\begin{frontmatter}

\title{A New Blackbody Radiation Law Based on Fractional Calculus and its Application to NASA COBE Data}

\author[1]{Minoru Biyajima}
\address[1]{Department of Physics, Shinshu University, Matsumoto 390-8621, Japan}
\ead{biyajima@azusa.shinshu-u.ac.jp}
\author[2]{Takuya Mizoguchi}
\address[2]{National Institute of Technology, Toba College, Toba 517-8501, Japan\corref{telephone/fax: +81599258088}}
\ead{mizoguti@toba-cmt.ac.jp}
\author[3]{Naomichi Suzuki}
\address[3]{Matsumoto University, Matsumoto 390-1295, Japan}
\ead{suzuki@matsu.ac.jp}

\begin{abstract}
By applying fractional calculus to the equation proposed by M. Planck in 1900, we obtain a new blackbody radiation law described by a Mittag--Leffler (ML) function. We have analyzed NASA COBE data by means of a non-extensive formula with a parameter $(q-1)$, a formula proposed by Ertik et al. with a fractional parameter $(\alpha-1)$, and our new formula including a parameter $(p-1)$, as well as the Bose--Einstein distribution with a dimensionless chemical potential $\mu$. It can be said that one role of the fractional parameter $(p-1)$ is almost the same as that of chemical potential $(\mu)$ as well as that of the parameter $(q-1)$ in the non-extensive approach.
\end{abstract}


\end{frontmatter}


\section{\label{sec1}Introduction}
The NASA COBE Collaboration has reported that the universe is full of photons at a temperature of 2.725 K \cite{Mather1994,Fixen1996,Fixsen2002,Nasa2005}. Their distribution is described by the Planck blackbody radiation law as follows:
\begin{eqnarray}
  U_{\rm Planck}(T,\,\nu) = \frac{C_B}{e^x - 1},
  \label{equ1} 
\end{eqnarray}
where $C_B = 8\pi h \nu^3/c^3$, $x=h\nu/k_BT$, $h$, $k_B$, and $T$ are the Planck's constant, the Boltzmann's constant, and temperature, respectively. $c$ is the speed of the light. 

Moreover, the following residual spectrum has been reported:
\begin{eqnarray*}
   {\rm [residual\ spectrum]} 
   ={\rm [COBE\ data] - [Eq.\ (\ref{equ1})\ with}\ T_{\rm CMB}\ {\rm K]}.
\end{eqnarray*}
It is worthwhile to notice that there are two kinds of residual spectra (1994) and (1996). To explain the residual spectrum mentioned above \cite{Mather1994,Fixen1996,Fixsen2002}, the following Bose--Einstein distribution with a dimensionless chemical potential $\mu$ is utilized by the NASA COBE Collaboration \cite{Sugiyama2001}:
\begin{eqnarray}
   U^{\rm (BE)}(T,\,\nu,\,\mu) = \frac{C_B}{e^{x+\mu} - 1}
   \ \mapright{\ |\mu| \ll 1\ }\  C_B\left[\frac 1{e^{x}-1} - \mu\frac{e^x}{(e^{x}-1)^2}\right].
  \label{equ2} 
\end{eqnarray}

To investigate the residual spectrum (1994) \cite{Mather1994,Fixen1996}, Tsallis et al. \cite{Tsallis1995} computed the following formula based on a non-extensive approach.
\begin{eqnarray}
   U^{\rm (NETD\ I)}(T,\,\nu,\,q) \cong  U_{\rm Planck}(T,\,\nu)
  + C_B\frac{q-1}{e^x-1}
   \times\left[ \ln(1-e^{-x}) - x\frac{1+e^{-x}}{1-e^{-x}}
  + \frac{x^2}2\frac{1+3e^{-x}}{(1-e^{-x})^2} \right],
  \label{equ3} 
\end{eqnarray}
where $(q-1)$ is known as the non-extensive parameter \cite{Tsallis1995,Plastino1995,Biyajima2012}. The correction term to the Planck distribution in Eq. (\ref{equ3}) is carefully calculated in \cite{Plastino1995}. The residual spectrum (1996) is investigated in \cite{Biyajima2012}.

Second, it is also known that Tirnakli et al. \cite{Tirnakli1998} have calculated the following formula based on the non-extensive approach, i.e., $q$ algebra (see Ref. \cite{Tsallis2009,Wilk2000,Beck2003,Biyajima2006}):
\begin{eqnarray}
   U^{\rm (NETD\ II)}(T,\,\nu,\,q) = \frac{C_B}{[1+(q-1)x]^{1/(q-1)}-1}
  \ \mapright{\ (q-1)\ll 1\ }\ 
  C_B\left[\frac 1{e^x-1} + \frac{q-1}2\frac{e^xx^2}{(e^x-1)^2}\right]. 
  \label{equ4} 
\end{eqnarray}

Third, using a generalized partition function derived from \cite{Haar1966,Wu1994,Suzuki1998}, Ertik et al. have proposed the following formula based on fractional calculus with the Mittag--Leffler (ML) function \cite{Ertik2009}:
\begin{eqnarray}
   U^{\rm (FC\ I)}(T,\,\nu,\,\alpha) = \frac{C_B}{E_{\alpha}(x) - 1}
  \mapright{\ (\alpha -1)\ll 1\ }\ 
  C_B\left[\frac{1}{e^x-1} + (\alpha -1)\frac{f(x)}{(e^x-1)^2}\right],
  \label{equ5} 
\end{eqnarray}
where the Mittag--Leffler (ML) function is defined as follows:
\begin{eqnarray*}
E_{\alpha}(x) = \sum_{n=0}^{\infty} \frac{x^n}{\Gamma(n\alpha + 1)}.
\end{eqnarray*}
The function $f(x)$ is expressed as follows:
\begin{eqnarray*}
f(x) = \sum_{k=0}^{\infty} \frac{kx^k\psi(1+k)}{\Gamma(1+k)}, 
\end{eqnarray*}
where $\psi(z) = d(\ln \Gamma (z))/dz$ is the digamma function, and $(\alpha -1)$ is known as the fractional parameter. Those distributions are summarized in Table \ref{tabl1}.

\begin{table*}[htbp]
  \centering
  \caption{\label{tabl1}Typical modified Planck distributions}
  \renewcommand{\arraystretch}{1.2}
  \begin{tabular}{c|c|l}
  \hline
  Bose--Einstein distributions    & $\dfrac 1{e^{x+\mu}-1}$ & $\mu$ : dimensionless chemical potential\\
  \hline
  \lw{Non-extensive approach II} & \lw{$\dfrac 1{e_q^x-1}$} & $q$ : non-extensive parameter\\
                                 &  & $e_q^x \equiv [1+(1-q)x]^{1/(1-q)}$\\
  \hline
  \lw{Fractional calculus I}     & \lw{$\dfrac 1{E_{\alpha}(x)-1}$} & $E_{\alpha}(x)$ : Mittag--Leffler function\\
                                 &  & $\alpha$ : fractional parameter\\
  \hline
  \end{tabular}
\end{table*}

In this study, we investigate the third approach, the fractional calculus, in more detail, because we are interested in the approach in \cite{Ertik2009}. Concerning the above-mentioned problem, we would like to adopt a different viewpoint from that of Ertik et al. \cite{Ertik2009}. It is well known that in the derivation of the blackbody radiation law in 1900, Planck adopted the following thermodynamical equation \cite{Planck2000a,Planck2000b,Sommerfeld1956} (See also \cite{Tsallis2009}):
\begin{eqnarray}
  \diff U{\beta} = - aU - bU^2,
  \label{equ6} 
\end{eqnarray}
where $U$ is the energy density. $\beta = 1/k_BT$ is the inverse temperature, and $a$ and $b$ are parameters. Using the ordinary calculus, we obtain the following expression:
\begin{eqnarray}
  U(x) = \frac{a/b}{e^x-1},  \quad x=a(\beta-\beta_0).
  \label{equ7} 
\end{eqnarray}

If we introduce a fractional derivative instead of the partial derivative for $\beta$ in Eq. (\ref{equ6}),  we cannot derive an analytical solution, because the equation is nonlinear for the function $U$. Therefore, we put $U=1/R$ and $x=a(\beta-\beta_0)$. Then, we obtain the following equation for $R$ from  Eq. (\ref{equ6}):
\begin{eqnarray}
  \diff R{x} =  R + \frac{b}{a}.
  \label{equ8} 
\end{eqnarray}
The solution for Eq. (\ref{equ8}) is given by
\begin{eqnarray}
  R(x) = \frac{1}{U(x)} = \frac{b}{a}\Bigl( {\rm e}^x - 1 \Bigr).
  \label{equ9} 
\end{eqnarray}
Equation (\ref{equ8}) is linear for $R$. Therefore, when the  Riemann--Liouville  fractional derivative is introduced in Eq. (\ref{equ8}) instead of the partial derivative of $x$, the Mittag--Leffler function appears in the solution \cite{Mittag-Leffler1905,Podlubny2009,West2003,Metzler2000,Suzuki2000}.

In this study, we aim to apply fractional calculus to Eq. (\ref{equ8}) in \S\ref{sec2}. In \S\ref{sec3}, various properties of the new formula in addition to Eqs. (\ref{equ2}), (\ref{equ4}), and (\ref{equ5}) are investigated. In \S\ref{sec4}, analysis of the NASA COBE data in terms of the new formula, as well as Eqs. (\ref{equ1}), (\ref{equ2}), (\ref{equ4}), and (\ref{equ5}) are presented. In \S\ref{sec5}, the concluding remarks and discussions are presented. 

In \ref{secA}, we explain an introduction of the fractional calculus in Eq. (\ref{equ8}). In \ref{secB}, Caputo derivative is mentioned. In \ref{secC}, an interrelation between chemical potential $\mu$ and fractional parameter $(p-1)$ is investigated. The same investigation in cases of NEXT II and FC I are briefly mentioned.


\section{\label{sec2}Application of Fractional Calculus to Eq. (\ref{equ8})}

The Riemann--Liouville fractional derivative \cite{Podlubny2009,West2003} of function $R(x)$ for $m=$1, 2, ... is defined as follows:
\begin{eqnarray}
  {}_0D^p_x R(x) = \frac{1}{\Gamma(m-p)}\Bigl(\frac{d}{dx}\Bigr)^m \int_0^x (x-\tau)^{m-p-1} R(\tau) d\tau
 \quad {\rm for}\ m-1 \le p < m.  
  \label{equ10} 
\end{eqnarray} 
From Eq. (\ref{equ10}), we have ${}_0D^p_x R(x) = \frac {d^{m-1}}{dx^{m-1}}R(x)$, if $p=m-1$. The Riemann--Liouville fractional integral is defined as follows:
\begin{eqnarray}
  {}_0D^{-p}_x R(x) = \frac{1}{\Gamma(p)} \int_0^x (x-\tau)^{p-1} R(\tau) d\tau
\quad {\rm for}\ 0 < p.
  \label{equ11} 
\end{eqnarray}

To obtain a fractional blackbody radiation formula or Bose--Einstein distribution, the partial derivative 
for $x$ in Eq. (\ref{equ8}) is replaced by the Riemann--Liouville fractional derivative (\ref{equ10}). 
Then, the following equation is obtained: 
\begin{eqnarray}
  {}_0D^p_x R(x) = R(x) + b/a,  \quad  x=a(\beta-\beta_0)>0.
  \label{equ12} 
\end{eqnarray}

If $p=1$, Eq. (\ref{equ12}) is reduced to a first order partial differential equation, Eq. (\ref{equ8}). Therefore, we seek the solution for $0<p<2$.

The Laplace transform $\tilde{R}(s)$ of function $R(x)$ is defined as follows:
\begin{eqnarray}
  \tilde{R}(s) = \mathcal{L}[R(x);s] = \int_0^\infty {\rm e}^{-sx} R(x) dx.
  \label{equ13} 
\end{eqnarray}

Using the following formulas,
\begin{eqnarray*}
  \mathcal{L}[{}_0D^p_x R(x);s] &\!\!\!=&\!\!\! s^p \tilde{R}(s) - \sum_{k=0}^{m-1} s^k\,\,{}_0D^{p-k-1}_x R(x)|_{x=0}
  = s^p \tilde{R}(s) - \sum_{k=0}^{m-1} c_k s^k, \\ 
  \mathcal{L}[1;s] &\!\!\!=&\!\!\! {1}/{s}, 
\end{eqnarray*}
where $c_k = \,{}_0D^{p-k-1}_x R(x)|_{x=0}$, we obtain from Eq. (\ref{equ12}) the Laplace transform $\tilde{R}(s)$,
\begin{eqnarray}
  \tilde{R}(s) =   \frac{b}{as(s^p - 1)} + \sum_{k=0}^{m-1} c_k \frac{s^k}{s^p-1}.
  \label{equ14} 
\end{eqnarray}

To obtain the solution $R(x)$ from Eq. (\ref{equ14}), we use the following formulas \cite{Podlubny2009} 
of the Laplace transform,
\begin{eqnarray}
  \hspace*{-5mm} \mathcal{L}[x^{\beta-1} E_{\alpha,\beta}( \gamma x^\alpha);s]
               = \frac{s^{\alpha-\beta}}{s^\alpha - \gamma }, \quad {\rm Re}(s) > |\gamma |^{1/\alpha},
  \label{equ15} 
\end{eqnarray}
where $ E_{\alpha,\beta}(x)$ denotes the generalized Mittag--Leffler (GML) function \cite{Podlubny2009,West2003} 
defined by
\begin{eqnarray}
  E_{\alpha,\beta}(x) \equiv \sum_{k=0}^{\infty} \frac{x^k}{\Gamma(\alpha k+ \beta)}\quad \alpha > 0,\ \beta >0.   
  \label{equ16} 
\end{eqnarray}

Then the solution $R(x)$ for $m-1\le p < m$ ($m=$1, 2, ...) is written as follows:
\begin{eqnarray}
  \hspace*{-5mm} R(x) = \frac{b}{a} x^p E_{p,p+1}(x^p) + \sum_{k=0}^{m-1} c_k x^{p-k-1} E_{p,p-k}(x^p).
  \label{equ17} 
\end{eqnarray}

It should be noticed that at $x=0$ the second term of the RHS in Eq. (\ref{equ17}) diverges, unless $c_0=0$ for $0<p<1$, and $c_1=0$ for $0<p<2$. If $p=1$, it is reasonable that Eq. (\ref{equ17}) reduces to Eq. (\ref{equ8}). Therefore, we assume that $c_0=c_1=0$, and we have
\begin{eqnarray}
  R(x) = \frac{b}{a} x^p E_{p,p+1}(x^p).
  \label{equ18} 
\end{eqnarray}

From Eqs. (\ref{equ16}) and (\ref{equ18}), we have
\begin{eqnarray*}
  R(x) = \frac{b}{a} \sum_{k=0}^\infty \frac{(x^p)^{k+1}}{\Gamma(p(k+1)+1)}
       = \frac{b}{a} \left\{ \sum_{k=0}^\infty \frac{(x^p)^k}{\Gamma(pk+1)} - 1\right\} 
       = \frac{b}{a} \left\{ E_{p}(x^p) - 1 \right\}.
\end{eqnarray*}
Therefore, Eq. (\ref{equ18}) coincides with Eq. (\ref{equ9}) for $p=1$. As for the constant $c_k$, using the following equation, 
\begin{eqnarray*}
  {}_0D^q_xR(x) &\!\!\!=&\!\!\! \frac{b}{a} \sum_{k=0}^\infty \frac{1}{\Gamma(p(k+1)+1)}  {}_0D^q_x x^{p(k+1)}
= \frac{b}{a} \sum_{k=0}^\infty \frac{x^{-q+p(k+1)}}{\Gamma(-q+p(k+1)+1)}
= \frac{b}{a} x^{p-q} E_{p,p-q+1}(x^p),
\end{eqnarray*}
which denotes the fractional derivative of $R(x)$ for $q>0$ and the fractional integral of $R(x)$ for $q<0$, we have
\begin{eqnarray*}
  c_k &\!\!\!=&\!\!\! \,{}_0D^{p-k-1}_x R(x)|_{x=0}= \frac{b}{a} x^{k+1} E_{p,k+2}(x^p)|_{x=0},
 \quad k=0,\ 1. 
\end{eqnarray*}
Therefore, our assumptions that $c_0=0$ for $0<p<1$ and $c_0=c_1=0$ for $1 \le p <2$ are satisfied.

Then the solution $R(x)$ of Eq. (\ref{equ18}) and a new form of Planck's black body radiation law are given respectively as follows:
\begin{eqnarray}
  R(x) &\!\!\!=&\!\!\! \frac{b}{a} \left\{ E_{p}(x^p) - 1 \right\},  \quad x=a(\beta-\beta_0), \nonumber\\
  U(x) &\!\!\!=&\!\!\! \frac{1}{R(x)} = \frac{a/b}{ E_{p}(x^p) - 1}.
  \label{equ19} 
\end{eqnarray}
Another possible approach to obtain the fractional blackbody radiation formula is briefly discussed in \ref{secB}. Using Eq. (\ref{equ19}), we can analyze NASA COBE data \cite{Nasa2005}. 


\section{\label{sec3}Various properties of Eqs. (\ref{equ2}), (\ref{equ4}), (\ref{equ5}), and (\ref{equ19})}

\begin{figure}
  \centering
  \includegraphics[width=76mm]{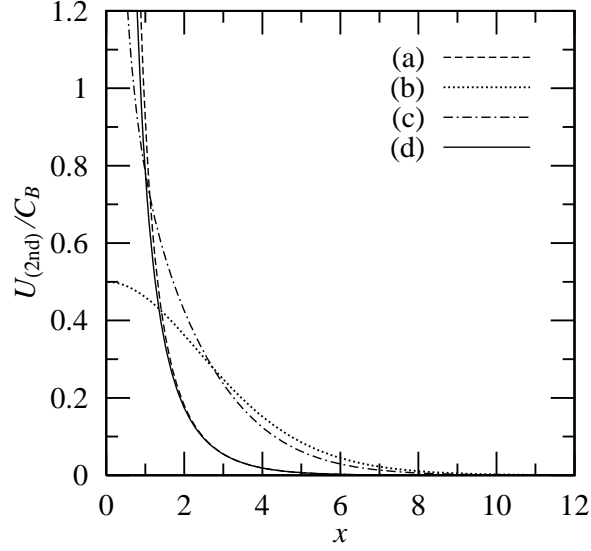}
  \caption{\label{fig1}Behavior of the second terms (named as $U_{\rm (2nd)}$) of the RHS's in Eqs. (\ref{equ2}), (\ref{equ4}), (\ref{equ5}), and (\ref{equ19}) without the coefficients factors $-\mu$, $(q-1)$, $(\alpha -1)$, and $(p-1)$. (a) Eq. (\ref{equ2}) (BE), (b) Eq. (\ref{equ4}) (NETD II), (c) Eq. (\ref{equ5}) (FC I), and (d) Eq. (\ref{equ19}) (FC II) with $\beta_0=0$.}
\end{figure}

We now investigate various properties of Eqs. (\ref{equ2}), (\ref{equ4}), (\ref{equ5}), and (\ref{equ19}). Concerning Eq. (\ref{equ19}), we have following approximate expression with $C_B$, $\beta_0=0$ and $x=h\nu/k_B T$, and named as (FC II).
\begin{eqnarray}
  {\rm Eq.\ (\ref{equ19})}:  U^{\rm (FC\ II)}(T,\,\nu,\,p) = \frac{C_B}{ E_p(x^p) - 1}
  \ \mapright{\ |p-1| \ll 1\ }\ 
   C_B\left[\frac 1{e^x-1} + \frac{p-1}{(e^x-1)^2}
  \sum_{k=0}^{\infty} \frac{kx^k[\psi(k+1)-\ln x]}{\Gamma(k+1)}\right].
  \label{equ20} 
\end{eqnarray}
Eq. (\ref{equ20}) is compared with Eq. (\ref{equ5}). The difference is seen in the second terms of the RHS's. The behaviors of the second terms (named as $U_{\rm (2nd)}$) of the RHS's in Eqs. (\ref{equ2}), (\ref{equ4}), (\ref{equ5}), and (\ref{equ20}) without the coefficients, $-\mu$, $(q-1)$, $(\alpha -1)$, and $(p-1)$ are presented in Fig. \ref{fig1}. This figure suggests that the Eq. (\ref{equ19}) is almost the same as the Bose--Einstein distribution.

Various analytic corrections for the Stephan--Boltzmann (SB) law without $C_B$ which is expressed as $U_{\rm (2nd)}x^3$ are calculated as follows:\medskip\\
Bose--Einstein distribution:
\begin{eqnarray}
  -\mu \int_0^{\infty}\frac{x^3\cdot e^x}{(e^x-1)^2}dx=-\mu \cdot 3 \cdot 2 \cdot \zeta(3),
  \label{equ21}  
\end{eqnarray}
where $\zeta(3)$ is the Riemann's $\zeta$ function.\medskip\\
Non-extensive formula II:
\begin{eqnarray}
   \frac 12(q-1)\int_0^{\infty}\frac{x^3\cdot e^xx^2}{(e^x-1)^2}dx
   = (q-1) \cdot \Gamma(6)\sum_{m=1}^{\infty} \left(\frac 1m\right)^5.
  \label{equ22} 
\end{eqnarray}
Fractional calculus I:
\begin{eqnarray}
   (\alpha-1)\int_0^{\infty}\frac{x^3}{(e^x-1)^2}\sum_{k=0}^{\infty} \frac{kx^k\psi(k+1)}{\Gamma(k+1)}dx
  = (\alpha-1) \sum_{k=0}^{\infty}\sum_{m=2}^{\infty} \frac{k(m-1)}{\Gamma(k+1)}\left(\frac 1m\right)^{k+4}\!\!\!\! \psi(k+1)\Gamma(k+4).
  \label{equ23} 
\end{eqnarray}
Fractional calculus II:
\begin{eqnarray}
  &&\hspace*{-13mm} (p-1)\int_0^{\infty}\frac{x^3}{(e^x-1)^2}\sum_{k=0}^{\infty} \frac{kx^k[\psi(k+1)-\ln x]}{\Gamma(k+1)}dx \nonumber\\
  &&\hspace*{-13mm} =(p-1) \sum_{k=0}^{\infty}\sum_{m=2}^{\infty} \frac{k(m-1)}{\Gamma(k+1)}\left(\frac 1m\right)^{k+4}\Gamma(k+4)
  \left[\ln m - \left(\frac 1{k+3}+\frac 1{k+2}+\frac 1{k+1}\right)\right].
  \label{equ24} 
\end{eqnarray}

Ref. \cite{Gradshteyn1965} is utilized in the calculations above. Numerically estimated values of the corrections to the modified SB law ($U_{(\rm 2nd)}/C_B$) are presented in Table \ref{tabl2}. It is seen that the roles of $\mu$ in Eq. (\ref{equ2}) and $(p-1)$ in Eq. (\ref{equ20}) are almost the same.

\begin{figure}
  \centering
  \includegraphics[width=76mm]{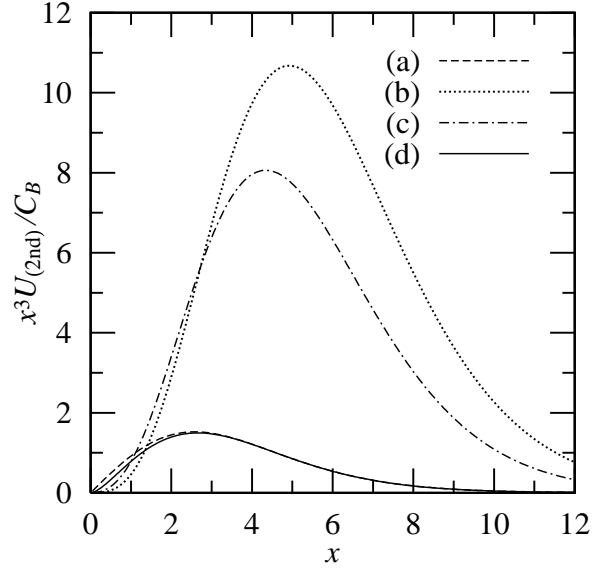}
  \caption{\label{fig2}Behavior of the integrands of Eqs. (\ref{equ21})--(\ref{equ24}) without the coefficients factors $-\mu$, $(q-1)$, $(\alpha -1)$, and $(p-1)$. (a) Eq. (\ref{equ21}) (BE), (b) Eq. (\ref{equ22}) (NETD II), (c) Eq. (\ref{equ23}) (FC I), and (d) Eq. (\ref{equ24}) (FC II).}
\end{figure}

\begin{table}[htbp]
  \centering
  \caption{\label{tabl2}Numerical coefficient factors of $\mu$, $(\alpha-1)$, $(q-1)$ and $(p-1)$ in Eqs. (\ref{equ21})--(\ref{equ24}).}
  \begin{tabular}{cl}
  \hline
  Bose--Einstein distribution & $-\mu \times 6\zeta(3) = -\mu \times 7.2123414$\\
  Non-extensive formula II & $(q-1)\times 62.22$\\
  Fractional calculus I    & $(\alpha-1)\times 44.41$\\
  (Ertik et al.)           &\\
  Fractional calculus II   & $(p-1)\times 6.9432884$\\
  \hline
  \end{tabular}
\end{table}


\section{\label{sec4}Analysis of NASA COBE data by Eqs. (\ref{equ1}), (\ref{equ2}), (\ref{equ4}), (\ref{equ5}), and (\ref{equ20})}

We are interested in COBE data and now analyze those data in terms of Eqs. (\ref{equ21})--(\ref{equ23}). It is known that the COBE data comprise distortion described by a very small dimensionless chemical potential ($\mu$) and/or the Sunyaev--Zeldovich (SZ) effect \cite{Biyajima2012}, and/or a possible effect called the distortion of the space-dimension \cite{Biyajima2014} (see the explanation in \S\ref{sec1}, Introduction.) The first terms of these distributions are the Planck distribution. The second terms are different with respect to each other. They correspond to the chemical potential ($\mu$) introduced by the NASA COBE Collaboration \cite{Mather1994,Fixen1996,Fixsen2002}.

\begin{figure}
  \centering
  \includegraphics[width=76mm]{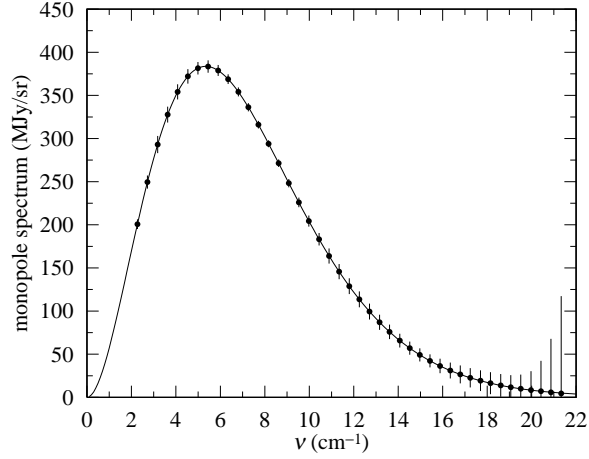}
  \caption{\label{fig3}Analysis of the NASA COBE monopole data in terms of Eq. (\ref{equ20}) (FC II): $T=2.7250$ K and $p-1 = 1.1\times 10^{-5}$ ($\chi^2/NDF = 45.0/41$). Error bars are 400$\sigma$. Data are taken from \cite{Nasa2005}.}
\end{figure}

In particular, our analysis is presented in Fig. \ref{fig3} and Table \ref{tabl3}. Combining the results in Tables \ref{tabl2} and \ref{tabl3}, we obtain the correction factors for the modified SB law $U/C_B$ for the COBE data. The ratios of the correction factors to those of the BE distribution are also shown in the parentheses of Table \ref{tabl4}.

\begin{table*}[htbp]
  \centering
  \caption{\label{tabl3}Analysis of NASA COBE monopole data in terms of Eqs. (\ref{equ1}), (\ref{equ2}), (\ref{equ4}), (\ref{equ5}), and (\ref{equ20}).}
  \begin{tabular}{cccc}
  \hline
  Eqs. & $T$ (K) & $\mu$, $(q-1)$, $(\alpha-1)$ or $(p-1)$ & $\chi^2$/NDF\\
  \hline
  Planck dis.           & $2.72502\pm 0.00001$ & ---                             & $45.1/42$\\
  BE: (\ref{equ2})      & $2.72501\pm 0.00002$ & $(-1.1\pm 3.2)\times 10^{-5}$   & $45.0/41$\\
  NETD II: (\ref{equ4}) & $2.72502\pm 0.00003$ & $(-0.53\pm 4.98)\times 10^{-6}$ & $45.1/41$\\
  FC I: (\ref{equ5})    & $2.72503\pm 0.00006$ & $(-0.22\pm 1.38)\times 10^{-5}$ & $45.1/41$\\
  FC II: (\ref{equ20})  & $2.72501\pm 0.00003$ & $(1.1\pm 3.5)\times 10^{-5}$    & $45.0/41$\\
  \hline
  \end{tabular}
\end{table*}

\begin{table*}[htbp]
  \centering
  \caption{\label{tabl4}Corrections to the modified SB law based on analysis of monopole COBE data in terms of Eqs. (\ref{equ1}), (\ref{equ2}), (\ref{equ4}), (\ref{equ5}), and (\ref{equ20}). Numerical factors in parentheses represent the ratios of the limits to that of the correction for the modified SB law.}
  \renewcommand{\arraystretch}{1.2}
  \begin{tabular}{rl}
  \hline
  Bose--Einstein dis.       & $-\mu \times 7.2123 = (7.87\pm 23.43)\times 10^{-5}$\\
                           & $|-\mu| \times 7.2123 <5.47\times 10^{-4}$\qquad (1.00)\\
  Non-extensive formula II & $(q-1)\times 62.22 = (-3.29\pm 31.01)\times 10^{-5}$\\
                           & $|q-1|\times 62.22 < 6.53\times 10^{-4}$\qquad (1.19)\\
  Fractional calculus I    & $(\alpha-1)\times 44.41 = (-9.51\pm 61.01)\times 10^{-5}$\\
  (Formula proposed by Ertik et al.) & $|\alpha-1|\times 44.41 < 1.32\times 10^{-3}$\qquad (2.41)\\
  Fractional calculus II   & $(p-1)\times 6.9432 = (7.39\pm 24.42)\times 10^{-5}$\\
                           & $|p-1| \times 6.9432 <5.62\times 10^{-4}$\qquad (1.03)\\
  \hline
  \end{tabular}
\end{table*}


\section{\label{sec5}Concluding remarks and discussions}
\noindent
{\bf C1)} By applying fractional calculus to the celebrated equation by M. Planck, i.e., Eq. (\ref{equ6}), we obtain a formula described by the ML function (Eq. (\ref{equ20})) which is different from Eq. (\ref{equ5}) proposed by Ertic et al. \cite{Ertik2009}. See \ref{secA} and \ref{secB}.\medskip\\
{\bf C2)} The behavior of Eq. (\ref{equ20}) is very similar to that of the Bose--Einstein distribution (see Fig. \ref{fig1} and Table \ref{tabl1}). 

In particular, from analysis of the COBE monopole data in terms of Eqs. (\ref{equ2}) and (\ref{equ20}), we obtain the following limits (see Table \ref{tabl3}):
\begin{eqnarray*}
  |\mu| &<&7.6\times 10^{-5} \quad ({\rm 95\ \% CL}),\\
  |p-1| &<&8.1\times 10^{-5} \quad ({\rm 95\ \% CL}).
\end{eqnarray*}

\begin{figure}
  \centering
  \includegraphics[width=76mm]{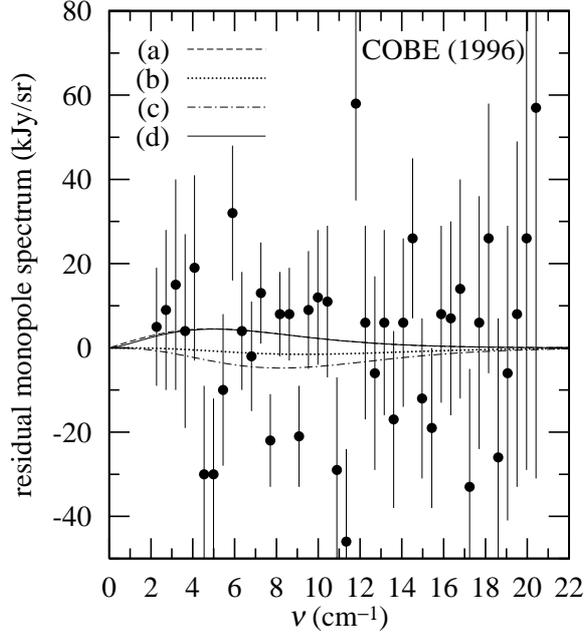}
  \caption{\label{fig4}Analysis of the residual monopole spectrum. (a) Eq. (\ref{equ2}) (BE) with $\mu = -1.1\times 10^{-5}$ ($\chi^2/NDF = 45.9/43$), (b) Eq. (\ref{equ4}) (NETD II) with $q-1 = -0.53\times 10^{-6}$ ($\chi^2/NDF = 45.2/43$), (c) Eq. (\ref{equ5}) (FC I) with $\alpha-1 = -0.22\times 10^{-5}$ ($\chi^2/NDF = 46.9/43$), and (d) Eq. (\ref{equ20}) (FC II) with $p-1 = 1.1\times 10^{-5}$ ($\chi^2/NDF = 45.9/43$).}
\end{figure}

As is seen in Fig. \ref{fig4} (residual spectrum (1996)) and Table \ref{tabl4}, it is difficult to distinguish among the Bose--Einstein distribution ($U^{\rm (BE)}$), Eq. (\ref{equ4}) ($U^{\rm (NEXT II)}$) based on $q$-algebra, and Eq. (\ref{equ20}) ($U^{\rm (FC II)}$) based on fractional calculus. These are able to describe the distortion of the COBE data.

The fractional calculus probably reflects the spectral distortion of CMB (Cosmic Microwave Background) of the ensemble (the universe) and contains the memory effect at the age of the universe. See also \ref{secB} and \ref{secC}.\medskip

\noindent
{\bf D1)} When we analyze the NASA COBE monopole data by means of Eq. (\ref{equ20}) (FC II) including $a\beta_0 = -\mu$, we have the following values: $T=2.72502\pm 4\times 10^{-5}$ K, $\mu = (-3.3\pm 5.8)\times 10^{-4}$, and $(p-1) = (-3.5\pm 6.3)\times 10^{-4}$. These figures depend on the initial values in the CERN MINUIT program. Estimated $\mu$ and $(p-1)$ are larger than those in Table \ref{tabl3}. This fact suggests that we have to choose one freedom between $a\beta_0 = -\mu$ and $(p-1)$: In the present study $a\beta_0 = 0$ is chosen. The reason of one freedom in FC II is presented in \ref{secC}: Analyzing NASA COBE data by Eq. (\ref{equ20}) including chemical potential $\mu$, we find the strong correlation between $\mu$ and fractional parameter $(p-1)$. This fact implies that we should choose one freedom between them.\medskip\\
{\bf D2)} Analyzing the same data by formula of NEXT II, Eq. (\ref{equ4}) with chemical potential $\mu$, we find the week correlation between them. See \ref{secC}. The magnitude of $|p-1| < 2.3 \times 10^{-5}$ in Table \ref{tabl5} seems to correspond to the SZ effect ($y$) reported by NASA COBE \cite{durrer2008}. (See Ref. \cite{Biyajima2012}.) As seen in Fig. \ref{fig6} b), the situation in FC I, i.e., the interrelation between the chemical potential $\mu$ and fractional parameter $(\alpha - 1)$ is intermediate among three cases.

\section*{Acknowledgements}
Authors would sincerely like to thank Prof. J. C. Mather and Prof. D. J. Fixsen for their kindness in communications for NASA COBE data and the method of analysis at an early stage of this investigation. One of the authors (M. B.) is grateful to Prof. N. Sugiyama for insightful conversations. Moreover, we sincerely would like to acknowledge Prof. M. Caputo for his kind sending of various papers, and Prof. Y. Sumino for his kind suggestions about geophysics.


\appendix

\section{\label{secA}Why do we adopt the fractional derivative in Eq. (\ref{equ8})?}
To describe an expansion of the universe at $\mu$-era after the Big Bang, Kompaneetz \cite{kompaneets1957} and Weymann \cite{weymann1965} propose the following equation for the photon distribution $f$,
\begin{eqnarray}
\diff ft = \frac{\kappa T_e}{m_ec^2}\frac{n_e\sigma_e}c x_e^{-2}\diff{}{x_e}x_e^4\left(\diff f{x_e} + f+ f^2\right)
  \label{equA1} 
\end{eqnarray}
where $T_e$ and $T$ are the electron and radiation temperatures, $n_e$ is the electron density, $\sigma_e$ is the Thomson cross-section and $x_e = h\nu/k_BT_e$. The stationary solution of Eq. (\ref{equA1}) is known as the Bose--Einstein distribution:
\begin{eqnarray}
f(x_e,\:\mu) = \frac 1{ce^{x_e} - 1} = \frac 1{e^{x_e+\mu} - 1},
  \label{equA2} 
\end{eqnarray}
where the chemical potential is introduced, reflecting that the number of photons are conserved due to the Compton scattering $\gamma + e^- \leftrightarrow \gamma + e^-$.

At 380k years after the Big Bang, the CMB (cosmic microwave background) photons are released. The CMB photons are described by the Planck distribution, because the number of photons are not conserved. Actually, the following processes occur in the universe: The double Compton and Bremsstrahlung scattering $\gamma + e^- \to e^- + 2\gamma$, $e^- + X \to e^- + X + \gamma$, where $X$ denotes an atomic nucleus, usually a proton or Helium-4 nucleus.

To understand the meaning of the chemical potential $\mu$ in Eq. (\ref{equA1}) before the age of recombination in a different point of view, we have to take into account the memory effect in Eq. (\ref{equ8}): Notice that the variable changes to $x_e \to x = h\nu/k_BT$, because the universe has cooled \cite{Metzler2000}. 

One of possible methods for taking into account the memory effect of the $\mu$-era is an introduction of fractional calculus. Following investigations in papers (or books) on geophysics by Caputo and/or Mainardi \cite{Caputo1967,Carpin1997}, we can replace the ordinary derivative by the fractional derivative in Eq. (\ref{equ6}), i.e., the celebrated equation by M. Planck in 1900.


\section{\label{secB}Comparison of the Riemann--Liouville fractional derivative and the Caputo fractional derivative}

In the present study, we use the Riemann--Liouville fractional derivative to extend Eq. (\ref{equ8}) to fractional order, and obtain the fractional blackbody radiation formula, Eq. (\ref{equ19}).  However, the possible initial condition to  Eq. (\ref{equ12}) is restricted : As is seen from Eq. (\ref{equ17}), if the initial condition $R(0)$ is not equal to zero, Eq. (\ref{equ12}) has no solution. 

If the Caputo fractional derivative~\cite{Caputo1967,Podlubny2009} is introduced into Eq. (\ref{equ12}) instead of the Riemann--Liouville fractional derivative, we can also obtain Eq. (\ref{equ19}) as a solution. Moreover, we can take the initial condition flexibly as in the ordinary differential equation. 
The Caputo fractional derivative of function $R(x)$  for $m=1$, $2$, ... is defined as
\begin{eqnarray}
  {}^C_0\! D^p_x R(x) = \frac{1}{\Gamma(m-p)} \int_0^x (x-\tau)^{m-p-1} R^{(m)}(\tau) d\tau,
  \quad m-1 < p < m,
\label{equB1} 
\end{eqnarray}
where $R^{(m)}(\tau) = \left(\frac{d}{d\tau}\right)^m R(\tau)$. From Eq. (\ref{equB1}), We obtain $\displaystyle{ \lim_{p\rightarrow m}{}^C_0\! D^p_x R(x) = R^{(m)}(x)}$. The Caputo fractional derivative is expressed by the use of the Riemann--Liouville fractional derivative and integral as
\begin{eqnarray}
 {}^C_0\! D^p_x R(x) =  {}_0\! D^{-\nu}_x ( {}_0\! D^m_x R(x) ), 
 \quad \nu = m -  p \ge 0.
\label{equB2} 
\end{eqnarray}
The difference between the Riemann--Liouville fractional derivative and the Caputo derivative appears distinctly in the differentiation of power function $x^\mu$. If $R(x)=x^\mu$ with $\mu=0$, $1$, ..., $m-1$, we obtain that ${}^C_0\! D^p_x\,x^\mu = 0$ in the Caputo fractional derivative. However, ${}_0\! D^p_x \,x^\mu = \{\Gamma(\mu+1)/\Gamma(-p+\mu+1)\}x^{-p+\mu}\neq 0$, in the Riemann--Liouville fractional derivative.

In the definition of the Riemann--Liouville fractional derivative, Eq. (\ref{equ10}), or the Caputo fractional derivative, Eq. (\ref{equB1}), integral from $x=0$ to $x$ is included. Therefore, the memory effect \cite{Caputo1967,Carpin1997} from the initial stage is taken into account in either of fractional derivatives .


\section{\label{secC}Interrelation between chemical potential $\mu$ and fractional parameter $(p-1)$, and the same studies for cases of NEXT II and FC I}
\begin{figure}
  \centering
  \includegraphics[width=76mm]{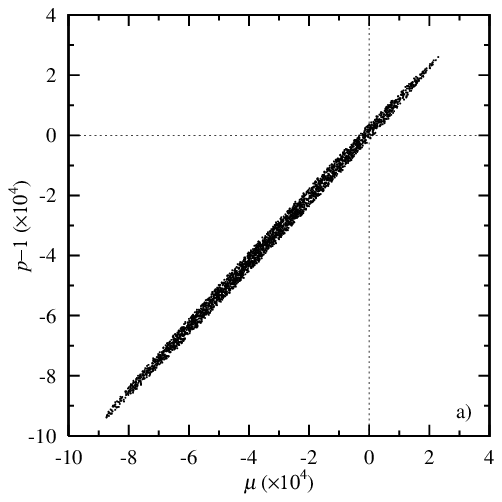}
  \includegraphics[width=76mm]{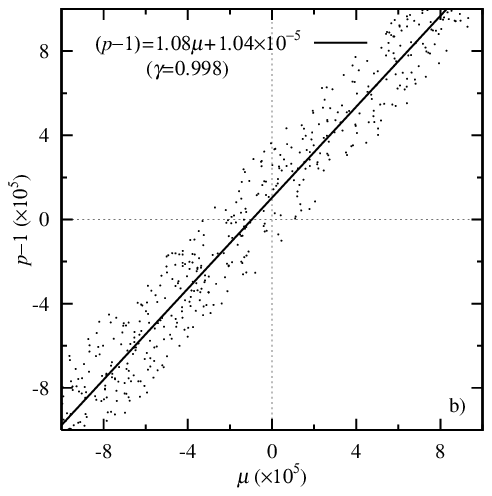}
  \caption{\label{fig5}Ensemble of parameters sets ($\mu$ and $(p-1)$) with constraint $\chi^2 < \chi_{min}^2$ (44.66 in Table \ref{tabl5}) $+ 1.0$. a) Number of satisfactory data 3021/ that of events generations is 500 k. b) Enlarged ensemble with Eq. (\ref{equC3}).}
\end{figure}

To elucidate roles of chemical potential $\mu$ and fractional parameter $(p-1)$, we adopt Eq. (\ref{equ19}) with $a\beta_0 = \mu$, 
\begin{eqnarray}
  U(x) = \frac{1}{ E_{p}(-(x-\mu)^p) - 1}.
  \label{equC1} 
\end{eqnarray}
Applying Eq. (\ref{equC1}) to the COBE data, we obtain results in Table \ref{tabl5}. To look for an interrelation between $\mu$ and $(p-1)$, we use a method of Monte Carlo calculus in analyses of COBE data: To understand contents of $\chi^2$-minimum by the CERN MINUIT in Table \ref{tabl5}, we adopt an allowed constraint, $\chi^2 \leq \chi_{\rm min}^2$(Table \ref{tabl5}) + 1.0 \cite{Biyajima2014}. Moreover, the following set of variables are prepared and 500k generations are performed: 
\begin{eqnarray}
  \left\{
  \begin{array}{l}
  T   = \langle T\rangle + \delta T\times {\rm Random\ number\ in\ the}\ [-1,\:1]\\
  \mu = \langle T\rangle + \delta \mu\times {\rm Random\ number\ in\ the}\ [-1,\:1]\\
  (p-1)=\langle p-1\rangle + \delta (p-1)\times {\rm Random\ number\ in\ the}\ [-1,\:1]
  \end{array}
  \right.
  \label{equC2} 
\end{eqnarray}
A number of satisfactory sets ($\chi^2 \leq \chi_{\rm min}^2 + 1.0$) is 3021/500k with $T = 2.72502\pm 0.00004$ K. This ensemble is shown in Fg. \ref{fig5}. 
\begin{table*}[htbp]
  \centering
  \caption{\label{tabl5}Analysis of COBE data by Eqs. (\ref{equ19}), (\ref{equ4}) and (\ref{equ5}) including $\beta_0$.}
  \begin{tabular}{ccccc}
  \hline
  Eq. & $T$ (K) & $\mu$ & $(p-1)$, $(q-1)$ or $(\alpha-1)$ & $\chi^2$/NDF\\
  \hline
  FC II: (\ref{equ19}) & $2.72502\pm 0.00004$ & $(-3.3\pm 5.7)\times 10^{-4}$  & $(-3.5\pm 6.1)\times 10^{-4}$ & $44.7/40$\\
                      & & $|\mu| < 1.5\times 10^{-3}$ (95\%\ CL)   & $|p-1| < 1.6\times 10^{-3}$ (95\%\ CL)\\
  \hline
  NETD II: (\ref{equ4}) & $2.72497 \pm 0.00010$ & $(-3.0 \pm 6.3)\times 10^{-5}$  & $(3.5 \pm 9.7)\times 10^{-6}$ & $44.9/40$\\
                      & & $|\mu| < 1.6\times 10^{-4}$ (95\%\ CL)   & $|q-1| < 2.3\times 10^{-5}$ (95\%\ CL)\\
  \hline
  FC I: (\ref{equ5}) & $2.72492 \pm 0.00022$ & $(-4.4\pm 8.7)\times 10^{-5}$  & $(1.5\pm 3.7)\times 10^{-5}$ & $44.8/40$\\
                      & & $|\mu|< 2.2\times 10^{-4}$ (95\%\ CL)   & $|\alpha-1|< 8.9\times 10^{-5}$ (95\%\ CL)\\
  \hline
  \end{tabular}
\end{table*}
To investigate an interrelation between $(p-1)$ and $\mu$, we use the method of linear regression. The following equation is obtained 
\begin{eqnarray}
  (p-1) = 1.08\,\mu + 1.04\times 10^{-5}
  \label{equC3} 
\end{eqnarray}
where the correlation coefficient $\gamma = 0.998$. From Eq. (\ref{equC3}), it can be said that $(p-1)$ and $\mu$ are strongly correlated. In other words, $(p-1)$ and $\mu$ are not independent each other: The role of the chemical potential $\mu$ is probably replaced by the fractional parameter $(p-1)$. Indeed this fact is seen in Table \ref{tabl4} (in the case of $a\beta_0 = \mu = 0$)
\begin{eqnarray*}
  |p-1| \leq 8.1\times 10^{-5}\ {\rm (95\%\: CL)}
\end{eqnarray*}
which can be compared with $|p-1| < 1.6\times 10^{-3}$ (95\% CL) and $|\mu| < 1.5\times 10^{-3}$ (95\% CL) in Table \ref{tabl5}. In other words, those magnitudes are larger than $|p-1| < 8.1\times 10^{-5}$. Thus since we cannot determine $\mu$ and $(p-1)$ simultaneously in Eq. (\ref{equC1}), we choose the fractional parameter $(p-1)$ between them.
\begin{figure}
  \centering
  \includegraphics[width=76mm]{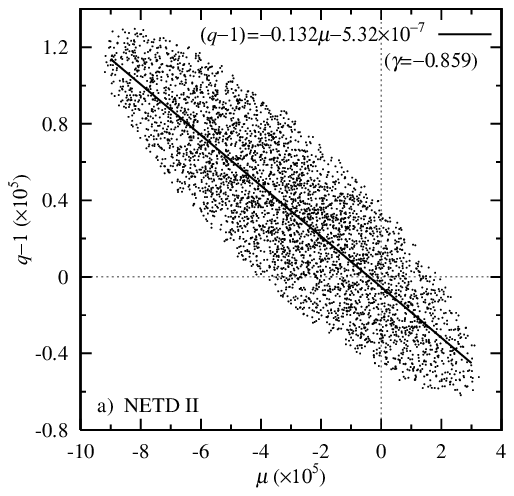}
  \includegraphics[width=76mm]{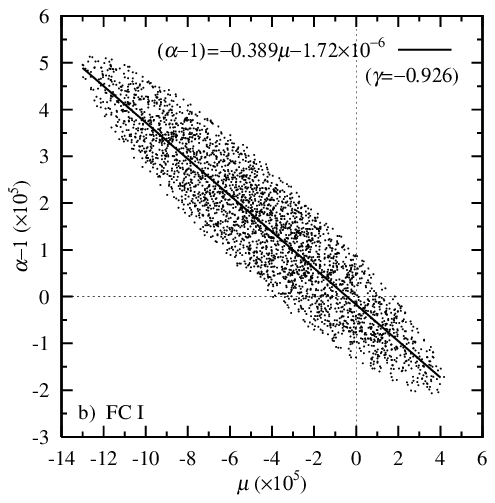}
  \caption{\label{fig6} a) Ensemble of parameters sets ($\mu$ and $(q-1)$) with constraint $\chi^2 < \chi_{min}^2$ (44.86) $+ 1.0$. Number of satisfactory data 4269/ that of event generation is 200 k. b) Ensemble of parameters sets ($\mu$ and $(\alpha-1)$) with constraint $\chi^2 < \chi_{min}^2$ (44.81) $+ 1.0$. Number of satisfactory data 3173/ that of event generation is 500 k.}
\end{figure}

By making use of the same method for the NEXT II and the FC I, we obtain results in Fig. \ref{fig6} and the following equations:
\begin{eqnarray}
  (q-1) &\!\!\!=&\!\!\! -0.132\, \mu - 5.32\times 10^{-7}\quad (\gamma = -0.859),
  \label{equC4}\\ 
  (\alpha-1) &\!\!\!=&\!\!\! -0.389\, \mu - 1.72\times 10^{-6}\quad (\gamma = -0.926).
  \label{equC5} 
\end{eqnarray}
In the case of NEXT II, the correlation between $(q-1)$ and $\mu$ seems to be week. They are probably independent quantities, on the contrary to the case of FC II. As seen in Table \ref{tabl5} and Fig. \ref{fig6} b), in the case of FC I, the situation of ($\mu$ and $(\alpha-1)$) is similar to the case of the NEXT II. 

In conclusion, it should be stressed that those facts above mentioned are based on analyses of COBE data.


\end{document}